\documentclass[12pt]{article}
\usepackage{amsfonts}
\usepackage[mathscr]{eucal}

\newcommand{\be}{\begin{equation}}
\newcommand{\ee}{\end{equation}}
\newcommand{\bea}{\begin{eqnarray}}
\newcommand{\eea}{\end{eqnarray}}
\newcommand{\ov}{\overline}
\newcommand{\ve}{\varepsilon}
\newcommand{\eps}{\epsilon}
\newcommand{\ba}{\begin{array}}
\newcommand{\ea}{\end{array}}
\newcommand{\cchi}{\raisebox{2pt}{$\chi$}}
\newcommand{\mb}{\mathbb}
\newcommand{\mf}{\mathfrak}
\newcommand{\ms}{\mathscr}

\begin{document}

\begin{flushright}
hep-th/9804107 \\
ITEP-98-11
\end{flushright}

\vspace{1.5cm}

\centerline{\LARGE\bf The top Yukawa coupling from 10D}
\centerline{\LARGE\bf supergravity with 
{\boldmath $E_8\times E_8$} matter}

\bigskip

\centerline{\Large K.Zyablyuk\footnote{\tt zyablyuk@heron.itep.ru}}

\bigskip

\centerline{\it Institute of Theoretical and Experimental Physics, Moscow}

\begin{abstract}
We consider the compactification of N=1, D=10 supergravity with 
$E_8\times E_8$ Yang-Mills matter to N=1, D=4 model with 3 generations.
With help of embedding $SU(5)\to SO(10)\to E_6\to E_8$ we find
the value of the top Yukawa coupling 
$\lambda_t = \sqrt{16\pi\alpha_{GUT}/3}$ at the GUT scale.
\end {abstract}

\section{Introduction}

Although superstring theories have great success and today are the 
best candidates for a quantum theory unifying all interactions,
they still do not predict any experimentally testable value,
mostly because there is no unambiguous procedure of compactification of
extra spatial dimensions.

On the other hand, many phenomenological models based on the unification
group $SO(10)$ were constructed (for instance \cite{so10}). They
describe quarks and leptons in representations $16_1$, $16_2$,
$16_3$ and Higgses, responsible for $SU(2)$ breaking, in $10$.
Basic assumption of these models is that there is only one 
Yukawa coupling $16_3\cdot 10\cdot 16_3$ at the GUT scale, which
gives masses of third generation. Masses of first two families and
mixings are generated due to the interaction with additional superheavy
($\sim M_{GUT}$) states in $16+\ov{16}$, $45$, $54$. Such models
explain generation mass hierarchy and allow to express all
Yukawa matrices, which well fit into experimentally observable pattern,
in terms of few unknown parameters.

Models of \cite{so10} are unlikely derivable from "more
fundamental" theory like supergravity/string. Nevertheless, 
something similar can
be constructed from D=10 supergravity coupled to $E_8\times E_8$
matter, which is low-energy limit of heterotic string
\cite{het}.

The lagrangian of N=1, D=10 supergravity with Yang-Mills matter
\cite{CM} does not contain any free parameter. The gauge group
is fixed to be either $SO(32)$ or $E_8\times E_8$ due to Green-Schwarz
anomaly cancellation mechanism \cite{GS}. We will consider only
$E_8\times E_8$ case, since it naturally leads to $SO(10)$ group.
Furthermore, since 
${\rm Tr}_{E_8^{(1)}\times E_8^{(2)}}=
{\rm Tr}_{E_8^{(1)}}+{\rm Tr}_{E_8^{(2)}}$,
fields from $E_8^{(2)}$ interact with $E_8^{(1)}$ only gravitationally,
so we will consider only $E_8^{(1)}$.

After compactification to four dimensions 10D-vector produces
6 real (or 3 complex) scalars in representation 248 each. In order
to break supersymmetry $N=4\to N=1$, some of them should get
$SU(3)$-valued VEVs \cite{CHSW}. So $E_8$ group is broken down
to $E_6$:
$$
248\,=\,(78,1)\,+\,(27,3)\,+\,(\ov{27},\ov{3})\,+\,(1,8)
$$
In this way we get 9 generations in 27 of $E_6$ with 9 mirror
generations $\ov{27}$. After $E_6$ is broken to $SO(10)$, we
get 12 pairs $16+\ov{16}$, 18 copies of 10 and 3 ones of 45. To compute
masses of all these states, one should evaluate some 
compactification scheme, for example Scherk-Schwarz compactification
\cite{SS}. We shall consider this question in future publications \cite{Z}.

Nevertheless, since in the models of \cite{so10} the top quark
gets the mass at the tree level, in our case it is possible
to find the top Yukawa coupling without going deep in the details
of compactification. It indicates the fact, that gauge and
Yukawa couplings are unified in higher-dimensional supersymmetric theories.
It is the subject of this paper.

In section 2 we perform the reduction of the N=1, D=10 supergravity
to N=1, D=4 theory and rewrite the result in conventional form in
terms of Kahler- and super-potential, carefully pointing out
all assumptions and conventions.

In section 3 we propose a way to choose 3 massless generations 
and compute the value of the top Yukawa coupling.

In Appendix we consecutively construct representations of $SO(10)$,
$E_6$ and $E_8$ groups.

\section{Reduction N=1, D=10 {\boldmath $\to$} N=1, D=4}

Bosonic lagrangian of $N=1$, $D=10$ supergravity with Yang-Mills
matter has the form \cite{CM}:
\be
\label{l10}
L^{(10)}\,=\,{1\over 4}\,R\,+\,{1\over 2}\,\phi^{;\,M}\phi_{;\,M}\,+\,
{1\over 12}\,e^{2\phi}H^{MNP}H_{MNP}\,+\,{1\over 4}\,e^\phi\,
{\rm Tr}\left(F^{MN}F_{MN}\right)
\ee
We use the following conventions for the field-strength tensor $F$
and 3-form $H$:
\bea
F_{MN} & = & 2\,\partial_{[M}A_{N]}\,-\,2\,A_{[M}A_{N]} \nonumber \\
\label{fh}
H_{MNP} & = & 3\,\partial_{[M}B_{NP]}\,+\,6\,{\rm Tr}\left(A_{[M}\partial_N
A_{P]}\,-\,{2\over 3}\,A_{[M}A_NA_{P]}\right)
\eea
In (\ref{l10}), (\ref{fh}) the vector field 
$A_M=A_M^{\mathbb A}T^{\mathbb A}$, $T^{\mathbb A}$ are antihermitean
generators (in this section the gauge group is arbitrary).
In principle we could write the gauge coupling constant before
the trace, but it can be removed by means of rescaling, so it's
not physical; actual value of the gauge coupling in four dimensions is
determined by the VEV of the dilaton field $\phi$.

The following index notations are used here:
$$
\ba{l|lll}
\mbox{space}& \mbox{dimension} & \mbox{flat} & \mbox{world}       \\
\hline
\mbox{initial:}& D=10 & A,B,C,\ldots & M,N,P,\ldots             \\
\mbox{our:}& D=4 & \alpha,\beta,\gamma,\ldots & \mu,\nu,\lambda,\ldots  \\
\mbox{internal:}& D=6 & a,b,c,\ldots & m,n,p,\ldots 
\ea
$$
The Minkowski metric is $\eta_{AB}=(+,-,\ldots,-)$.

As usual, instead of 6 real coordinates $y^m$ of internal manifold
we introduce 3 complex ones:
\be
z^s\,=\,y^s\,+\,i\,y^{s+3} \;,\qquad
z^{\bar{s}}\,=\,y^s\,-\,i\,y^{s+3} \;.
\ee
We will use letters $s,t,u,...=1,2,3$ for complex indices.
If the supersymmetry transformation of the
vector  potential $\delta A_{\bar{s}} \sim \eps$, then 
$\delta A_s \sim \bar{\eps}$. So if $A_{\bar{s}}$ are supersymmetric
partners of left-handed fermions, then $A_s$ are partners of
right-handed (or mirror) ones.

\begin{table}
\begin{center}
\unitlength=1mm
\begin{picture}(165,70)(-19,-15)
\put(-18,31){\small graviton:}
\put(-18,24){\small gravitino:}
\put(-18,17){\small 2-form:}
\put(-18,10){\small dilatino:}
\put(-18,3){\small dilaton:}
\put(0,31){$E_M{}^A(35)$}
\put(2,24){$\psi_M^I(56)$}
\put(0,17){$B_{MN}(28)$}
\put(4,10){$\cchi^I(8)$}
\put(5,3){$\phi(1)$}
\put(18,31){$=$}
\put(18,24){$=$}
\put(18,17){$=$}
\put(18,10){$=$}
\put(18,3){$=$}
\put(24,31){$E_\mu{}^\alpha(2)$}
\put(25,24){$\psi_\mu^4(2)$}
\put(24,17){$B_{\mu\nu}(1)$}
\put(26,10){$\cchi^4(2)$}
\put(27,3){$\phi(1)$}
\put(23,1){\line(0,1){19.5}}
\put(23,22){\line(0,1){13}}
\put(37,1){\line(0,1){19.5}}
\put(37,22){\line(0,1){13}}
\put(23,1){\line(1,0){14}}
\put(23,22){\line(1,0){14}}
\put(23,20.5){\line(1,0){14}}
\put(23,35){\line(1,0){14}}
\put(53,31){$+$}
\put(39,24){$+$}
\put(39,17){$+$}
\put(39,10){$+$}
\put(46.5,24){$\psi_\mu^i(3\times 2)$}
\put(45,17){$B_{\mu m}(6\times 2)$}
\put(47,10){$\cchi^i(3\times 2)$}
\put(44,8){\line(0,1){20.5}}
\put(66,8){\line(0,1){20.5}}
\put(44,8){\line(1,0){22}}
\put(44,28.5){\line(1,0){22}}
\put(68,24){$+$}
\put(74,31){$E_\mu{}^a (6\times 2)$}
\put(75,24){$\psi_m^4 (6\times 2)$}
\put(82,17){$+$}
\put(73,22){\line(0,1){13}}
\put(94,22){\line(0,1){13}}
\put(73,22){\line(1,0){21}}
\put(73,35){\line(1,0){21}}
\put(96,31){$+$}
\put(96,24){$+$}
\put(104,31){$E_s{}^i(9)$}
\put(102,24){$\psi_s^i(9\times 2)$}
\put(104,17){$B_{s\bar{t}}(9)$}
\put(101,15){\line(0,1){20}}
\put(119.5,15){\line(0,1){20}}
\put(101,15){\line(1,0){18.5}}
\put(101,35){\line(1,0){18.5}}
\put(121.5,31){$+$}
\put(121.5,24){$+$}
\put(121.5,17){$+$}
\put(129,31){$E_{\bar{s}}{}^i(12)$}
\put(127.5,24){$\psi_{\bar{s}}^i(9\times 2)$}
\put(129.5,17){$B_{st}(6)$}
\put(126.5,15){\line(0,1){20}}
\put(145,15){\line(0,1){20}}
\put(126.5,15){\line(1,0){18.5}}
\put(126.5,35){\line(1,0){18.5}}
\put(15,-9){\small dilaton chiral}
\put(19,-14){\small multiplet}
\put(30,-5){\vector(0,1){4}}
\put(44,-3){\small Part of N=4, D=4}
\put(44,-8){\small SUGRA multiplet}
\put(54,1){\vector(0,1){5}}
\put(98,3){\small chiral multiplets}
\put(96,-2){\small of moduli fields $T_{s\bar{t}}$}
\put(110,7){\vector(0,1){5}}
\put(20,48){\small N=1, D=4}
\put(15,43){\small SUGRA multiplet}
\put(30,41){\vector(0,-1){4}}
\put(71,48){\small $SO(6)$ vector}
\put(75,43){\small multiplet}
\put(83,41){\vector(0,-1){4}}
\end{picture}
\end{center}
\caption{N=1, D=10 SURGA multiplet at the reduction to four dimensions.
Index $I=1...4$ is a part of 10D spinorial index; it splits on
$(i, 4)$, $i=1,2,3$ after supersymmetry breaking $N=4 \to N=1$.
Numbers in brackets are physical degrees of freedom, carried by 
each field.}
\end{table}

Table 1 demonstrates, how the fields of 10D supergravity form
4D-multiplets at the reduction.  Not all of them can be coupled to
$N=1$, $D=4$ supergravity. At first, one should vanish the
part of $N=4$ multiplet, namely, 3 gravitinos $\psi^i_\mu$,
3 fermions $\cchi^i$ from dilatino and 6 vectors $B_{\mu m}$.
Since 2-form $B$ is not gauge invariant, 
$\delta B_{MN}=-2{\rm Tr}(U\partial_{[M}A_{N]})$ at the gauge 
transformations $\delta A_M=D_MU$, the condition $B_{\mu m}=0$
does not break the gauge invariance only if vectors $A_\mu$ and
scalars $A_m$ belong to different representations, so that
${\rm Tr}(T(A_\mu)T(A_m))=0$. This excludes the possibility to
couple adjoint scalars with N=1, D=4 supergravity in 
supersymmetric way. 

The fields in the very right box of the Table 1 mix left- and
right-handed generations. We do not know, is it possible
to couple them to N=1, D=4 SUGRA, so we put
$E_{\bar{s}}{}^i=B_{st}=\psi_{\bar{s}}^i=0$.
Again, these conditions are supersymmetric-invariant 
if we consider only left- or right-handed chiral fields $A_{\bar{s}}$,
but not both of them, so that ${\rm Tr}(T(A_{\bar{s}})T(A_{\bar{t}}))=0$.
It indicates, that reasons, responsible for $N=4\to N=1$ supersymmetry
breaking, also break the mirror symmetry.

We shall not consider vector multiplet $(E_\mu{}^a, \psi_m^4)$,
since $SU(3)$-holonomy group, gauged by these vectors, is broken,
and they become massive.

The reduction of other fields is quite standard \cite{FKP}.
As usual, 4-dimensional theory contains the following bosons:
$$
\ba{lccl}
\mbox{4D vielbein:} & e_\mu{}^\alpha & = & \Delta^{1/4} E_\mu{}^\alpha \\
\mbox{4D dilaton:} & S & = & \sqrt{\Delta} e^\phi\,-\,2iD \\
\mbox{moduli:} & T_{s\bar{t}} & = & -\,e^{-\phi} g_{s\bar{t}}\,-\,2\,
B_{s\bar{t}}\,-\,2\,{\rm Tr}(A_s A_{\bar{t}})
\ea
$$
where $\Delta=\det(g_{mn})$, $D$ is dual to $B_{\mu\nu}$,
$e^{2\phi}\Delta H_{\alpha\beta\gamma} = \ve_{\alpha\beta\gamma}{}^\delta
D_{;\,\delta}$. In these terms the bosonic part of 4-dimensional
lagrangian gets the form
\be
\label{l4n1}
L^{(4)}\,=\,{1\over 4}\,R\,
+\,{1\over 4}\,{\rm Tr}\left[\,({\rm Re}\,S)\,
F^{\alpha\beta}F_{\alpha\beta}\,-\,{1\over 2}\,({\rm Im}\,S)\,
\ve^{\alpha\beta\gamma\delta}
F_{\alpha\beta}F_{\gamma\delta}\,\right] +\,L^{(K)}\,-\,V
\ee
with kinetic terms in conventional Kahler form 
$L^{(K)}=G^{\rm i}{}_{\rm j}
D_\mu\ov{\Phi}_{\rm i}D^\mu\Phi^{\rm j}$, where
$\Phi^{\rm i}=(S, T_{s\bar{t}}, A_{\bar{s}})$, $G$ is Kahler potential
(see below). The potential in (\ref{l4n1}):
\bea
-V & = & {1\over {\rm Re}S\det(-g_{x\bar{y}})}\left\{
-{1\over 2}g_{w\bar{z}}\ve^{wsu}\ve^{ztv}
{\rm Tr}\!\left([A_s,A_u][A_{\bar{t}},A_{\bar{v}}]\right) - \right.
\nonumber\\     
\label{pot0}
 & & \left.-\,{8\over 3}\,{\rm Tr}(A_{[s}A_t A_{u]})
{\rm Tr}(A_{\bar{s}}A_{\bar{t}} A_{\bar{u}})\right\} -
\,{1\over 2\,{\rm Re}\,S}\,{\rm Tr}\!\left(
[A^{\bar{s}},A_{\bar{s}}][A^{\bar{t}},A_{\bar{t}}]\right)
\eea
It can be written in conventional form:
\be
\label{potcon}
-V\,=\,e^{2G}\left[\,{3\over 2}\,-\,(G^{\rm i}{}_{\rm j})^{-1}
G^{\rm j}G_{\rm i}\,\right] +
{1\over 2}\,({\rm Re}\,f_{\cal AB})^{-1}
\left[G_{\rm i} (T_{\cal A})^{\rm i}{}_{\rm j}\Phi^{\rm j}\right]\!
\left[G_{\rm k} (T_{\cal B})^{\rm k}{}_{\rm l}\Phi^{\rm l}\right]
\ee
with kinetic function $f_{\cal AB}=-S \cal{G}_{\cal AB}$,
$\cal{G}_{\cal AB}$ is Killing tensor which stands in the trace
${\rm Tr}(AB)={\cal G}_{\cal AB} A^{\cal A} B^{\cal B}$,
and Kahler potential:
\be
\label{kahp}
G\,=\,-\,{1\over 2}\,\ln{(\,S\,+\,\ov{S}\,)}\,-\,{1\over 2}\,
\ln{\det{\left[\,T_{s\bar{t}}\,+\,\ov{T}_{s\bar{t}}\,+\,4\,
{\rm Tr}(A_s A_{\bar{t}})\,\right]}}\,+\,{1\over 2}\,\ln{|W|^2}
\ee
$W$ is the superpotential:
\be
\label{spot1}
W\,=\,{8\sqrt{2}\over 3}\,\ve^{stu}\,{\rm Tr}(A_{\bar{s}}
A_{\bar{t}}A_{\bar{u}})
\ee
We use the convention for gravitational constant $4\pi G_N=1$.
In (\ref{potcon}) indices ${\cal A,B}$ label representation,
to which vectors belong. 

All fermionic terms can be restored unambiguously  by functions
$f_{\cal AB}$ and $G$.

\section{Top Yukawa coupling}

Now we consider low energy approximation to the lagrangian 
(\ref{l4n1}), (\ref{pot0}) with
group $E_8$. The structural constants of $E_8$ are given in
appendix (A.3); the trace convention is 
${\rm Tr}(AB)={1\over 30}{\mb G}_{\mb{AB}}A^{\mb A}B^{\mb B}$
with Killing tensor ${\mb G}_{\mb{AB}}$ from (\ref{e8kil})
(actually, as we mentioned earlier, this convention doesn't
matter unless we know the value of the dilaton $S$).

Suppose that the dilaton $S$ and metric $g_{s\bar{t}}$ get
some VEVs and consider the interactions of 9 copies of 27
$A^{i\mf a}_{\bar{s}}$, ${\mf a}=1...27$. We put:
\be
A^{i\mf a}_s\,=\,0 \;, \qquad
A^{i\mf a}_{\bar{s}}\,=\,{1\over\sqrt{2}}\,
e^{-\phi/2}e_{\bar{s}}{}^{\bar{j}}C^{i\mf a}_j
\ee
($A^{i\mf a}_s$ correspond to mirror generations $\ov{27}$;
$e_{\bar{s}}{}^{\bar{j}}$ is 6D vielbein in complex coordinates
$g_{s\bar{t}}=e_s{}^i\eta_{i\bar{j}}e_{\bar{t}}{}^{\bar{j}}$,
Minkowski metric is $\eta_{i\bar{j}}=-{1\over 2}\delta_{ij}$.)
Discarding nonrenormalizable interactions, one can write down the
lagrangian in the form
\be
\label{lle6}
L\,=\,-\,{{\rm Re}\,S\over 4}\,F^{\mf A}_{\mu\nu}F^{\mf{A}\,\mu\nu}\,+
\,(D_\mu C^{i\mf a}_j)^*D^\mu C^{i\mf a}_j\,-\,
\left({\partial W'\over \partial C^{i\mf a}_j}\right)^*
{\partial W'\over \partial C^{i\mf a}_j}\,+\,\mbox{$D$-terms}
\ee
with the superpotential
\be
\label{spot2}
W' \, = \, {1\over 6\sqrt{{\rm Re}\,S}}\,\ve^{ijk}\ve_{lmn}
d_{\mf{abc}}C^{l\mf a}_iC^{m\mf b}_jC^{n\mf c}_k \,= 
\, {1\over \sqrt{{\rm Re}\,S}}\,d_{\mf{abc}}\,\det
\left| \ba{ccc}
C^{1\mf a}_1 & C^{2\mf b}_1 & C^{3\mf c}_1 \\
C^{1\mf a}_2 & C^{2\mf b}_2 & C^{3\mf c}_2 \\
C^{1\mf a}_3 & C^{2\mf b}_3 & C^{3\mf c}_3 
\ea \right|
\ee
(it differs from (\ref{spot1}) in normalization).
Totally symmetric $E_6$-invariant tensor $d_{\mf{abc}}$
is determined in Appendix (\ref{dinv},\ref{dnorm});
index ${\mf A}$=1...78. $E_6$-covariant derivative in (\ref{lle6}) is:
\be
D_\mu C^{i\mf a}_j\,=\,\partial_\mu C^{i\mf a}_j\,-\,
A_\mu^{\mf A}(T_{\mf A})^{\mf a}{}_{\mf b}C^{i\mf b}_j \;,
\ee
normalization of generators 
${\rm Tr}(T_{\mf A}T_{\mf B})=-3\delta_{\mf{AB}}$, like in Appendix (A.3).

How to choose 3 generations from 9 ones in (\ref{spot2})? 
As mentioned in the Introduction, if we want to build a model
like \cite{so10}, we should have only the interaction 
$16_3\cdot 10\cdot 16_3$ in the superpotential at the GUT scale.
This term follows from $27_3\cdot 27_3\cdot 27_3$, if Higgs' 10
is the part of $27_3$. If we choose
\be
\label{mych}
C^{i\mf a}_j\,=\,\left(\ba{ccc}
{1\over\sqrt{3}}C^{\mf a}_3 & {1\over\sqrt{2}}C^{\mf a}_2 & C^{\mf a}_1 \\
0 & {1\over\sqrt{3}}C^{\mf a}_3 & {1\over\sqrt{2}}C^{\mf a}_2 \\
0 & 0 & {1\over\sqrt{3}}C^{\mf a}_3 \ea \right)
\ee
then we get the superpotential with only third generation:
\be
\label{spot3}
W'\,=\,{1\over 3\sqrt{3\,{\rm Re}\,S}}\,d_{\mf{abc}}\,
C^{\mf a}_3 C^{\mf b}_3 C^{\mf c}_3
\ee
Of course the choice (\ref{mych}) is not unique, other variants
leading to (\ref{spot3}) are possible (although they give the
same numerical factor, which determines the value of the top Yukawa
coupling, discussed here). Strictly speaking, one should find
the metric of 6D space, which would give masses to all states
except those of (\ref{mych}). We shall present such solution
in future publications \cite{Z}. Nevertheless the superpotential
does not depend on the moduli fields $T_{s\bar{t}}$, so we just accept
the choice (\ref{mych}).

Simple reduction, considered here, does not determine
the value of the dilaton ${\rm Re}\,S$. But it gives the relation
between the top Yukawa coupling and the gauge constant,
which is the evidence of the gauge--Yukawa unification.
To establish the correspondence between $\lambda_t$ and $\alpha_{GUT}$,
we perform two steps down $E_6\to SO(10)\to SU(5)$ by means of 
representations, constructed in Appendix.

{\boldmath $E_6\to SO(10)$}. The multiplet $27_3$
consists of $C_3^{\mf a}=(1,H^{\ms A},\Psi^{\rm a})$,
where $\Psi^{\rm a}$ are third quarks and leptons packed in 16
(a=1...16) and $H^{\ms A}$ are Higgses in 10 (${\ms A}$=1...10);
we shall not consider singlets here. Concerning vector fields,
we choose representation 45 from 78 one: 
$A_\mu^{\mf{A}=\ms{AB}}=\sqrt{2}A_\mu^{\ms{AB}}$. To determine the
value of the gauge and Yukawa couplings, we write down the following
3 elements of the theory:
\begin{itemize}
\item
Kinetic terms:
\be
-\,{{\rm Re}\,S\over 4}\,F^{\ms{AB}}_{\mu\nu}F^{\ms{AB}\,\mu\nu}\,+
(D_\mu\Psi^{\rm a})^*D^\mu\Psi^{\rm a}\,+
(D_\mu H^{\ms A})^*D^\mu H^{\ms A}
\ee
\item
Covariant derivatives:
\be
D_\mu\Psi^{\rm a}\,=\,\partial_\mu\Psi^{\rm a}\,-\,{1\over 4}\,
A_\mu^{\ms{AB}}(\Gamma^{\ms{AB}})^{\rm a}{}_{\rm b}\Psi^{\rm b}\;,\qquad
D_\mu H^{\ms A}\,=\,\partial_\mu H^{\ms A}\,-\,A_\mu^{\ms{AB}}H^{\ms B}
\ee
\item
The superpotential:
\be
W'\,=\,{1\over \sqrt{6\,{\rm Re}\,S}}\,\Psi^{\rm a}(\Gamma_{\ms A})_{\rm ab}
\Psi^{\rm b} H^{\ms A}
\ee
\end{itemize}

{\boldmath $SO(10)\to SU(5)$}. Now indices ${\sf a,b,c...}$ run
1...5 while ${\sf A,B,C}...=1...24$. As usual, we will use 24 hermitean
traceless $5\times 5$ matrices $(\lambda^{\sf A})^{\sf b}{}_{\sf c}$
normalized ${\rm Tr}(\lambda^{\sf A}\lambda^{\sf B})=2\delta^{\sf AB}$.
$SO(10)$-fields produce the following $SU(5)$ ones:
$$
\ba{lllll}
SO(10)\;\mbox{field:}& \mbox{IR:} & 
SU(5)\;\mbox{field:}& \mbox{IR:} & \mbox{particles:} \\ 
\Psi^{\rm a} & 16 & \psi^{\sf ab}=\Psi^{{\rm a}={\sf ab}} & 10 &
u_L,u_R,d_L,e_R \\
 & & \varphi_{\sf a}=\Psi^{{\rm a}={\sf a}} &\ov{5} & d_R,e_L,\nu_L \\
H^{\ms A} & 10 & H_u^{\sf a}={1\over\sqrt{2}}(H^{\sf a}+iH^{{\sf a}+5}) &
5 & \mbox{up-Higgs} \\
 & & H_{d\,{\sf a}}={1\over\sqrt{2}}(H^{\sf a}-iH^{{\sf a}+5}) & \ov{5}&
\mbox{down-Higgs} \\
A_\mu^{\ms{AB}} & 45 & A_\mu^{\sf A}=-{i\over 2}
(\lambda^{\sf A})^{\sf c}{}_{\sf b}(A_\mu^{\sf bc}-iA_\mu^{\sf b,c+5}+ & & \\
 & & \qquad\;\; +iA_\mu^{\sf b+5,c}+A_\mu^{\sf b+5,c+5}) & 24 & 
SU(5)\;\mbox{vector}
\ea
$$
We shall not consider singlet $\Psi^{{\rm a}=1}$ and other vectors,
not relevant to the $SU(5)$-unification model.

Elements of the theory:
\begin{itemize}
\item
Kinetic terms:
\be
\label{su5k}
-{{\rm Re}\,S\over 4}F^{\sf A}_{\mu\nu}F^{{\sf A}\,\mu\nu}+
{1\over 2}(D_\mu\psi^{\sf ab})^*D^\mu\psi^{\sf ab}+
(D_\mu\varphi_{\sf a})^*D^\mu\varphi_{\sf a}+
(D_\mu H_u^{\sf a})^*D^\mu H_u^{\sf a}+
(D_\mu H_{d\,\sf a})^*D^\mu H_{d\,\sf a}
\ee
\item
Covariant derivatives:
$$
D_\mu\psi^{\sf ab}\,=\,\partial_\mu\psi^{\sf ab}\,-\,
{i\over 2}\,A_\mu^{\sf A}(\lambda^{\sf A})^{\sf a}{}_{\sf c}\psi^{\sf cb}\,
-\,{i\over 2}\,A_\mu^{\sf A}(\lambda^{\sf A})^{\sf b}{}_{\sf c}\psi^{\sf ac}
\;,\quad
D_\mu\varphi_{\sf a}\,=\,\partial_\mu\varphi_{\sf a}\,+\,
{i\over 2}\,A_\mu^{\sf A}\varphi_{\sf c}(\lambda^{\sf A})^{\sf c}{}_{\sf a}
$$
\be
\label{su5d}
D_\mu H_u^{\sf a}\,=\,\partial_\mu H_u^{\sf a}\,-\,
{i\over 2}\,A_\mu^{\sf A}(\lambda^{\sf A})^{\sf a}{}_{\sf b}H_u^{\sf b}\;,
\qquad
D_\mu H_{d\,\sf a}\,=\,\partial_\mu H_{d\,\sf a}\,+\,
{i\over 2}\,A_\mu^{\sf A}H_{d\,\sf b}(\lambda^{\sf A})^{\sf b}{}_{\sf a}
\ee
\item
The superpotential:
\be
\label{su5s}
W'\,=\,{1\over \sqrt{3\,{\rm Re}\,S}}\left(\,2\,H_{d\,\sf a}\psi^{\sf ab}
\varphi_{\sf b}\,+\,{1\over 4}\,\ve_{\sf abcdf}H_u^{\sf a}
\psi^{\sf bc} \psi^{\sf df}\,\right)
\ee
\end{itemize}

The embedding of the Standard Model group in $SU(5)$ is well known.
From (\ref{su5k}), (\ref{su5d}) we immediately find the gauge
constant $g$ and from (\ref{su5s}) the top Yukawa coupling $\lambda_t$:
\be
g\,=\,{1\over\sqrt{{\rm Re}\,S}}\;,\qquad
\lambda_t\,=\,{2\over\sqrt{3\,{\rm Re}\,S}}
\ee
Eliminating unknown ${\rm Re}S$, we get the equation, which is
in principle experimentally testable:
\be
\label{result}
\lambda_t\,=\,\sqrt{16\pi\alpha_{GUT}\over 3}
\ee
where $\alpha_{GUT}=g^2/4\pi$. This constant determines
the value of the top quark mass at low energies (or at least 
its upper limit, since $\tan{\beta}$ is unknown):
\be
m_t\,=\,\lambda_t\,{v\over \sqrt{2}}\,\sin{\beta} \;,\qquad
v\,=\,{2\,m_Z\sin{\theta_W}\cos{\theta_W} \over e}\approx 246\,GeV
\ee

In order to evaluate the renormalization running of $\lambda_t$
from the GUT scale down to weak scale, one needs to know all
massless states of the theory in this range and their interactions. 
If we suppose, that all symmetries except $SU(3)\times SU(2)\times U(1)$
are broken somewhere near the GUT scale, and all massless states
are only those of MSSM, then $\lambda_t$ runs as Figure 1 shows.
In this case small $\tan{\beta}$ are excluded regardless of their
incompatibility with $b-t$ unification.

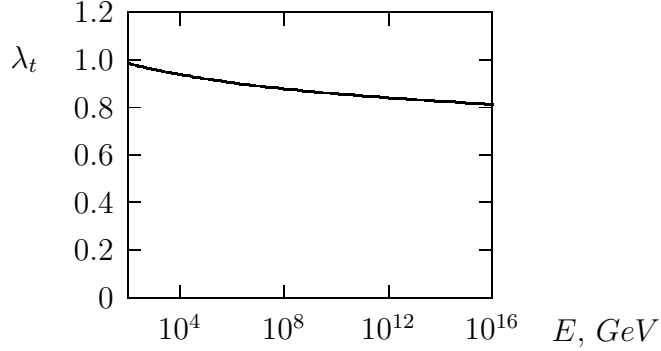
\begin{figure}
\begin{center}
\setlength{\unitlength}{0.240900pt}
\ifx\plotpoint\undefined\newsavebox{\plotpoint}\fi
\sbox{\plotpoint}{\rule[-0.175pt]{0.350pt}{0.350pt}}%
\begin{picture}(900,620)(100,100)
\put(930,90){$E,\,GeV$}
\put(80,520){$\lambda_t$}
\sbox{\plotpoint}{\rule[-0.175pt]{0.350pt}{0.350pt}}%
\put(264,158){\rule[-0.175pt]{137.795pt}{0.350pt}}
\put(264,158){\rule[-0.175pt]{4.818pt}{0.350pt}}
\put(242,158){\makebox(0,0)[r]{0}}
\put(816,158){\rule[-0.175pt]{4.818pt}{0.350pt}}
\put(264,233){\rule[-0.175pt]{4.818pt}{0.350pt}}
\put(242,233){\makebox(0,0)[r]{0.2}}
\put(816,233){\rule[-0.175pt]{4.818pt}{0.350pt}}
\put(264,308){\rule[-0.175pt]{4.818pt}{0.350pt}}
\put(242,308){\makebox(0,0)[r]{0.4}}
\put(816,308){\rule[-0.175pt]{4.818pt}{0.350pt}}
\put(264,383){\rule[-0.175pt]{4.818pt}{0.350pt}}
\put(242,383){\makebox(0,0)[r]{0.6}}
\put(816,383){\rule[-0.175pt]{4.818pt}{0.350pt}}
\put(264,457){\rule[-0.175pt]{4.818pt}{0.350pt}}
\put(242,457){\makebox(0,0)[r]{0.8}}
\put(816,457){\rule[-0.175pt]{4.818pt}{0.350pt}}
\put(264,532){\rule[-0.175pt]{4.818pt}{0.350pt}}
\put(242,532){\makebox(0,0)[r]{1.0}}
\put(816,532){\rule[-0.175pt]{4.818pt}{0.350pt}}
\put(264,607){\rule[-0.175pt]{4.818pt}{0.350pt}}
\put(242,607){\makebox(0,0)[r]{1.2}}
\put(816,607){\rule[-0.175pt]{4.818pt}{0.350pt}}
\put(346,158){\rule[-0.175pt]{0.350pt}{4.818pt}}
\put(346,113){\makebox(0,0){$10^4$}}
\put(346,587){\rule[-0.175pt]{0.350pt}{4.818pt}}
\put(509,158){\rule[-0.175pt]{0.350pt}{4.818pt}}
\put(509,113){\makebox(0,0){$10^8$}}
\put(509,587){\rule[-0.175pt]{0.350pt}{4.818pt}}
\put(673,158){\rule[-0.175pt]{0.350pt}{4.818pt}}
\put(673,113){\makebox(0,0){$10^{12}$}}
\put(673,587){\rule[-0.175pt]{0.350pt}{4.818pt}}
\put(836,158){\rule[-0.175pt]{0.350pt}{4.818pt}}
\put(836,113){\makebox(0,0){$10^{16}$}}
\put(836,587){\rule[-0.175pt]{0.350pt}{4.818pt}}
\put(264,158){\rule[-0.175pt]{137.795pt}{0.350pt}}
\put(836,158){\rule[-0.175pt]{0.350pt}{108.164pt}}
\put(264,607){\rule[-0.175pt]{137.795pt}{0.350pt}}
\put(264,158){\rule[-0.175pt]{0.350pt}{108.164pt}}
\sbox{\plotpoint}{\rule[-0.350pt]{0.700pt}{0.700pt}}%
\put(836,462){\usebox{\plotpoint}}
\put(824,462){\rule[-0.350pt]{2.891pt}{0.700pt}}
\put(807,463){\rule[-0.350pt]{4.095pt}{0.700pt}}
\put(795,464){\rule[-0.350pt]{2.891pt}{0.700pt}}
\put(779,465){\rule[-0.350pt]{3.854pt}{0.700pt}}
\put(762,466){\rule[-0.350pt]{4.095pt}{0.700pt}}
\put(746,467){\rule[-0.350pt]{3.854pt}{0.700pt}}
\put(730,468){\rule[-0.350pt]{3.854pt}{0.700pt}}
\put(718,469){\rule[-0.350pt]{2.891pt}{0.700pt}}
\put(701,470){\rule[-0.350pt]{4.095pt}{0.700pt}}
\put(689,471){\rule[-0.350pt]{2.891pt}{0.700pt}}
\put(673,472){\rule[-0.350pt]{3.854pt}{0.700pt}}
\put(660,473){\rule[-0.350pt]{3.132pt}{0.700pt}}
\put(644,474){\rule[-0.350pt]{3.854pt}{0.700pt}}
\put(632,475){\rule[-0.350pt]{2.891pt}{0.700pt}}
\put(619,476){\rule[-0.350pt]{3.132pt}{0.700pt}}
\put(607,477){\rule[-0.350pt]{2.891pt}{0.700pt}}
\put(595,478){\rule[-0.350pt]{2.891pt}{0.700pt}}
\put(583,479){\rule[-0.350pt]{2.891pt}{0.700pt}}
\put(570,480){\rule[-0.350pt]{3.132pt}{0.700pt}}
\put(558,481){\rule[-0.350pt]{2.891pt}{0.700pt}}
\put(550,482){\rule[-0.350pt]{1.927pt}{0.700pt}}
\put(538,483){\rule[-0.350pt]{2.891pt}{0.700pt}}
\put(530,484){\rule[-0.350pt]{1.927pt}{0.700pt}}
\put(517,485){\rule[-0.350pt]{3.132pt}{0.700pt}}
\put(509,486){\rule[-0.350pt]{1.927pt}{0.700pt}}
\put(501,487){\rule[-0.350pt]{1.927pt}{0.700pt}}
\put(489,488){\rule[-0.350pt]{2.891pt}{0.700pt}}
\put(481,489){\rule[-0.350pt]{1.927pt}{0.700pt}}
\put(472,490){\rule[-0.350pt]{2.168pt}{0.700pt}}
\put(464,491){\rule[-0.350pt]{1.927pt}{0.700pt}}
\put(456,492){\rule[-0.350pt]{1.927pt}{0.700pt}}
\put(448,493){\rule[-0.350pt]{1.927pt}{0.700pt}}
\put(440,494){\rule[-0.350pt]{1.927pt}{0.700pt}}
\put(432,495){\rule[-0.350pt]{1.927pt}{0.700pt}}
\put(427,496){\rule[-0.350pt]{1.204pt}{0.700pt}}
\put(419,497){\rule[-0.350pt]{1.927pt}{0.700pt}}
\put(411,498){\rule[-0.350pt]{1.927pt}{0.700pt}}
\put(403,499){\rule[-0.350pt]{1.927pt}{0.700pt}}
\put(399,500){\rule[-0.350pt]{0.964pt}{0.700pt}}
\put(391,501){\rule[-0.350pt]{1.927pt}{0.700pt}}
\put(382,502){\rule[-0.350pt]{2.168pt}{0.700pt}}
\put(378,503){\rule[-0.350pt]{0.964pt}{0.700pt}}
\put(370,504){\rule[-0.350pt]{1.927pt}{0.700pt}}
\put(366,505){\rule[-0.350pt]{0.964pt}{0.700pt}}
\put(362,506){\rule[-0.350pt]{0.964pt}{0.700pt}}
\put(354,507){\rule[-0.350pt]{1.927pt}{0.700pt}}
\put(350,508){\rule[-0.350pt]{0.964pt}{0.700pt}}
\put(342,509){\rule[-0.350pt]{1.927pt}{0.700pt}}
\put(338,510){\rule[-0.350pt]{0.964pt}{0.700pt}}
\put(333,511){\rule[-0.350pt]{1.204pt}{0.700pt}}
\put(325,512){\rule[-0.350pt]{1.927pt}{0.700pt}}
\put(321,513){\rule[-0.350pt]{0.964pt}{0.700pt}}
\put(317,514){\rule[-0.350pt]{0.964pt}{0.700pt}}
\put(313,515){\rule[-0.350pt]{0.964pt}{0.700pt}}
\put(309,516){\rule[-0.350pt]{0.964pt}{0.700pt}}
\put(301,517){\rule[-0.350pt]{1.927pt}{0.700pt}}
\put(297,518){\rule[-0.350pt]{0.964pt}{0.700pt}}
\put(293,519){\rule[-0.350pt]{0.964pt}{0.700pt}}
\put(289,520){\rule[-0.350pt]{0.964pt}{0.700pt}}
\put(284,521){\rule[-0.350pt]{1.204pt}{0.700pt}}
\put(280,522){\rule[-0.350pt]{0.964pt}{0.700pt}}
\put(276,523){\rule[-0.350pt]{0.964pt}{0.700pt}}
\put(272,524){\rule[-0.350pt]{0.964pt}{0.700pt}}
\put(268,525){\rule[-0.350pt]{0.964pt}{0.700pt}}
\put(264,526){\rule[-0.350pt]{0.964pt}{0.700pt}}
\end{picture}
\end{center}
\caption{The running of the top Yukawa coupling in case of MSSM
and $\alpha_{GUT}=1/25$.}
\end{figure}

Nevertheless, MSSM might be incomplete in all range up to the
GUT scale. Indeed, initially we had 9 generations $27$ and
9 mirrors $\ov{27}$. It is possible to find a configuration, 
which would give masses to all extra scalars and
make 6 massive Dirac fermions in representation $27$ of $E_6$.
Nevertheless, if 3 fermions in 27 are massless,
so do 3 mirror fermions in $\ov{27}$. Whatever the compactification is,
it seems difficult to make these 3 mirror generations massive,
unless $SU(2)$ is broken, which happens at the weak scale.
It would change the renormalization running.

\section{Conclusion}

Even though the main result of this paper (\ref{result})
is based on several assumptions which might be incorrect,
we demonstrated the predictive power of 10D supergravity
coupled to $E_8\times E_8$ matter. In order to verify them and
to get other predictions, one should: 1) consider a particular
anzatz for the metric of 6D-space $g_{mn}$, for instance, within the 
framework of Sherk-Schwarz compactification procedure and 2)
explore flat directions of the potential. Then masses of superheavy
states are expressed in terms of few parameters of 6D-space
and VEVs of Higgses, responsible for symmetry breaking. 
Masses of first  two families are generated due to interaction
with superheavy states.

Moreover, since higher derivative corrections to the action of
10D supergravity became recently available \cite{STZ},
the VEV of the dilaton can be expressed in terms of
the sizes of 6D space. It cannot be determined in another
calculable way within the framework of second-derivative supergravity
due to the dilaton runaway problem. It will be done in future
publications \cite{Z}.

\appendix 

\section{Appendix: groups {\boldmath $SO(10)$, $E_6$, $E_8$}}

For convenience we collected all group indices in Table 2.

\begin{table}[b]
$$
\ba{lll}
\hline
\mbox{indices:} \qquad & \mbox{values:} \qquad & \mbox{representation:} \\
\hline\hline
{\sf a,b,c,...}  & 1...5 & SU(5) \ \ \mbox{fundamental}  \\
{\sf A,B,C,...}  & 1...24 & SU(5) \ \ \mbox{adjoint}  \\
{\rm a,b,c,...}  & 1...16 & SO(10) \ \ \mbox{spinorial}  \\
\ms{A,B,C,...}  & 1...10 & SO(10) \ \ \mbox{vector} \\
\mf{a,b,c,...}  & 1...27 & E_6 \ \ \mbox{fundamental} \\
\mf{A,B,C,...}  & 1...78 & E_6 \ \ \mbox{adjoint} \\
i,j,k,...  & 1...3 & SU(3) \ \ \mbox{fundamental} \\
\Sigma,\Lambda,\Pi,...  & 1...8 & SU(3) \ \ \mbox{adjoint} \\
\mb{A,B,C,...}  & 1...248 & E_8 \ \ \mbox{adjoint} \\
\hline
\ea
$$
\caption{Group indices, their values and representations, which they label}
\end{table}

\subsection{{\boldmath $SO(10) \supset SU(5)\times U(1)$}}

Vector and spinorial representations of $SO(10)$ are decomposed
by $SU(5)$ reps:
$$
10\,=\,5\,+\,\ov{5}\;,\qquad
16\,=\,10\,+\,\ov{5}\,+\,1
$$
Let us denote $SO(10)$-vector index $\ms{A}=({\sf a},5+{\sf a})$=1...10,
${\sf a}=1...5$ and $SO(10)$-spinorial one 
${\rm a}=(1,{\sf a}, {\sf bc})=1...16$ with ${\sf b}<{\sf c}$.

We can construct symmetric $16\times 16$ Dirac matrices
$(\Gamma^{\ms A})_{\rm ab}$ and
$(\Gamma^{\ms A})^{\rm ab}=((\Gamma^{\ms A})_{\rm ab})^*$ in
euclidean 10D space in the following way:
\be
\label{g10f5}
(\Gamma^{\sf a})_{\rm bb'}\,=\left(\ba{ccc}
0 & \delta^{\sf a}_{\sf b'} & 0 \\
\delta^{\sf a}_{\sf b} & 0 & 2\delta^{\sf ab}_{\sf c'd'} \\
0 & 2\delta^{\sf ab'}_{\sf cd} & \ve_{\sf acdc'd'} \ea\right) \;,\qquad
(\Gamma^{5+{\sf a}})_{\rm bb'}\,=\,i\left(\ba{ccc}
0 & \delta^{\sf a}_{\sf b'} & 0 \\
\delta^{\sf a}_{\sf b} & 0 & -2\delta^{\sf ab}_{\sf c'd'} \\
0 & -2\delta^{\sf ab'}_{\sf cd} & \ve_{\sf acdc'd'} \ea\right) 
\ee
where ${\rm b}=(1,{\sf b},{\sf cd})$, ${\sf c}<{\sf d}$. Note, that 
one should sum over index 16 by the following
$$
\psi_{\rm b}\psi^{\rm b}\,=\,1\cdot 1\,+\,5_{\sf b} 5^{\sf b}\,+\,
\sum_{{\sf c}<{\sf d}}10_{\sf cd} 10^{\sf cd}
$$
to avoid double count, since $10^{\sf cd}$ is antisymmetric. 
$\Gamma$-matrices (\ref{g10f5}) satisfy anticommutational relations:
$$
(\Gamma^{\ms A})_{\rm ac} (\Gamma^{\ms B})^{\rm cb}\,+\, 
(\Gamma^{\ms B})_{\rm ac} (\Gamma^{\ms A})^{\rm cb}\,=\,
2\,\delta^{\ms{AB}}\delta_{\rm a}^{\rm b}
$$
with $\delta$-symbol 
$\delta_{\rm b'}^{\rm b}=(1,\,\delta^{\sf b}_{\sf b'},\, 
\delta^{\sf c}_{\sf c'}\delta^{\sf d}_{\sf d'})$.

\subsection{{\boldmath $E_6 \supset SO(10)\times U(1)$}} 

Now with help of these $\Gamma$-matrices
we construct generators of the $E_6$ group using the decomposition
$E_6 \supset SO(10) \times U(1)$:
$$
27\,=\,1\,+\,10\,+16\;,\qquad 78\,=\,1\,+\,16\,+\,\ov{16}\,+\,45
$$
Let us denote 27 index as $\mf{a}=(1,\ms{A}, {\rm a})$.
Consider the following $27\times 27$ antihermitean
traceless matrix:
\be
\label{e6gen}
T^{\mf{a}}{}_{\mf{b}}\,=\,
\left(\ba{ccc} is & 0 & \sqrt{2}\psi_{\rm b} \\
0 & J^{\ms A}{}_{\ms B}-{i\over 2}s\delta^{\ms A}_{\ms B}& 
(\bar{\psi}\Gamma^{\ms A})_{\rm b} \\
-\sqrt{2}\bar{\psi}{}^{\rm a} & -(\Gamma_{\ms B}\psi)^{\rm a} &
{1\over 4}J^{\ms{CD}}(\Gamma_{\ms{CD}})^{\rm a}{}_{\rm b}
+{i\over 4}s\delta^{\rm a}_{\rm b} \ea\right)
\ee
Here $s$ is real number, $J_{\ms AB}$ is antisymmetric $10\times 10$
matrix and $\psi_{\rm a}$ is 16-component complex spinor,
$\bar{\psi}{}^{\rm a}\equiv (\psi_{\rm a})^*$. The commutator
of both matrices of type (\ref{e6gen}) is also a matrix of this type:
$$
\left[T_{(1)},\,T_{(2)}\right]\,=\,T_{(3)}
$$
constructed from elements:
\bea
s_{(3)} & = & 
\,4\,{\rm Im}\left(\bar{\psi}_{(1)}\psi_{(2)}\right) \nonumber \\
J_{(3)}{}^{\ms A}{}_{\ms B} & = & [J_{(1)},\,J_{(2)}]{}^{\ms A}{}_{\ms B}\,
-\,\bar{\psi}_{(1)}\Gamma^{\ms A}{}_{\ms B}\psi_{(2)}\,
+\,\bar{\psi}_{(2)}\Gamma^{\ms A}{}_{\ms B}\psi_{(1)} \nonumber \\  
\label{e6com}
\psi_{(3)} & = & {1\over 4}\left(\,\hat{J}_{(1)}\psi_{(2)}
\,-\,\hat{J}_{(2)}\psi_{(1)}\,+\,3i\,s_{(1)}\psi_{(2)}
\,-\,3i\,s_{(2)}\psi_{(1)}\,\right)
\eea
where $\hat{J}=J_{\ms{AB}}\Gamma^{\ms{AB}}$.
This proves the algebraic structure of matrices (\ref{e6gen}).
Trace of both such matrices is:
\be
\label{e6tr}
{\rm Tr}(T_{(1)}T_{(2)})\,=\,-\,{9\over 2}\,s_{(1)}s_{(2)}\,
+\,3\,{\rm Tr}(J_{(1)}J_{(2)})\,
-\,12\,\bar{\psi}_{(1)}\psi_{(2)}\,-\,12\,\bar{\psi}_{(2)}\psi_{(1)}
\ee
Now let us enumerate 78 independent matrices (\ref{e6gen})
by index ${\mf A}$, 
$T_{\mf A}=(s,\psi_{\rm a}, \bar{\psi}{}^{\rm a}, J^{\ms{AB}})$.
$E_6$-metric $g_{\mf{AB}}\equiv {\rm Tr}(T_{\mf A}T_{\mf B})$ 
can be found by formula (\ref{e6tr}).

The $E_6$-structural constants
\be
\label{e6scd}
\left[T_{\mf A},\,T_{\mf B}\right]\,=
\,f^{\mf C}{}_{\mf{AB}}T_{\mf C}
\ee
can be found by (\ref{e6com}) with help of
$f^{\mf C}{}_{\mf AB}=g^{\mf{CD}}{\rm Tr}(T_{\mf D}
[T_{\mf A},\,T_{\mf B}])$, where $g^{\mf{AB}}$ is inverse
to $g_{\mf{AB}}$. 
With help of these expressions one can find the trace in adjoint 
representation:
\be
f^{\mf C}{}_{\mf{AD}}f^{\mf D}{}_{\mf{BC}}\,=\,4\,g_{\mf{AB}}
\ee
There exists totally symmetric invariant tensor $d_{\mf{abc}}$ in $E_6$. The
invariance condition is:
\be
\label{dinv}
d_{\mf{d}(\mf{ab}}T^{\mf d}{}_{\mf c)}\,=\,0
\ee
In the representation (\ref{e6gen}) all nonzero components of
this tensor are:
\be
d_{1\ms{AB}}\,=\,\delta_{\ms{AB}}\;,\qquad
d_{\ms{A}{\rm bc}}\,=\,{1\over \sqrt{2}}(\Gamma_{\ms A})_{\rm bc}
\ee
It satisfies the normalization condition:
\be
\label{dnorm}
d_{\mf{acd}}d^{\mf{bcd}}\,=\,10\,\delta^{\mf b}_{\mf a}\;,\qquad
d^{\mf{abc}}\equiv (d_{\mf{abc}})^*
\ee
The following Fiertz identity can be proved:
\be
g^{\mf{AB}}(T_{\mf A})^{\mf c}{}_{\mf a}(T_{\mf A})^{\mf d}{}_{\mf b}\,=\,
{1\over 18}\,\delta^{\mf c}_{\mf a}\delta^{\mf d}_{\mf b}\,+\,
{1\over 6}\,\delta^{\mf d}_{\mf a}\delta^{\mf c}_{\mf b}\,-\,
{1\over 6}\,d^{\mf{cdf}}d_{\mf{fab}}
\ee
Note, that Fiertz identities of all groups (except $E_8$) are given
in \cite{C}.

\subsection{{\boldmath $E_8\supset E_6 \times SU(3)$}}

Fundamental representation of $E_8$ coincides with adjoint one,
so we only need to find the structural constants with help of
decomposition $E_8\supset E_6\times SU(3)$:
$$
248\,=\,(78,1)\,+\,(27,3)\,+\,(\ov{27},\ov{3})\,+\,(1,8)
$$

Let us choose $E_6$-generators normalized as
${\rm Tr}(T_{\mf A}T_{\mf B})=-3\delta_{\mf{AB}}$.
The $SU(3)$ generators are 8 traceless hermitean
$3\times 3$ Gell-Mann matrices $(\lambda_\Sigma)^i{}_j$, 
$\Sigma=1...8$, $i,j=1,2,3$:
$$
[\lambda_\Sigma,\,\lambda_\Lambda]\,=\,2i\,c^\Pi{}_{\Sigma\Lambda}
\lambda_\Pi\;,\qquad
{\rm Tr}(\lambda_\Sigma\lambda_\Lambda)\,=\,2\,\delta_{\Sigma\Lambda}\;,
\qquad
c^{\Xi}{}_{\Sigma\Pi}c^\Pi{}_{\Lambda\Xi}\,=\,-\,3\,\delta_{\Sigma\Lambda}
$$

Generators of $E_8$ are 
${\mb T}_{\mb A}=(X_{\mf A}, Y_\Sigma, Q_{i \mf a}, Q^{i\mf a})$, where
$X_{\mf A}$, $Y_\Sigma$ are $E_6$ and 
$SU(3)$ generators, 
$Q_{i\mf a}$ and $Q^{i\mf a}\equiv (Q_{i\mf a})^+$. The
$E_8$ commutational relations, closed with respect to Jacobi identities, are:
$$
[X_{\mf A},\,X_{\mf B}]\,=\,f^{\mf C}{}_{\mf{AB}}X_{\mf C} \qquad
[Y_\Sigma,\,Y_\Lambda]\,=\,c^\Pi{}_{\Sigma\Lambda}Y_\Pi \qquad
[X_{\mf A},\,Y_\Sigma]\,=\,0
$$
$$
[X_{\mf A},\,Q_{i\mf a}]\,=\,(T_{\mf A})^{\mf b}{}_{\mf a} Q_{i\mf b} \qquad
[X_{\mf A},\,Q^{i\mf a}]\,=\,-\,(T_{\mf A})^{\mf a}{}_{\mf b} Q^{i\mf b}
$$
$$
[Y_\Sigma,\,Q_{i\mf a}]\,=\,-\,{i\over 2}\,(\lambda_\Sigma)^j{}_i
Q_{j\mf a} \qquad
[Y_\Sigma,\,Q^{i\mf a}]\,=\,{i\over 2}\,(\lambda_\Sigma)^i{}_j
Q^{j\mf a}
$$
$$
[Q_{i\mf a},\,Q_{j\mf b}]\,=\,{1\over \sqrt{2}}
\ve_{ijk}d_{\mf{abc}}Q^{k\mf c}
\qquad
[Q^{i\mf a},\,Q^{j\mf b}]\,=\,-\,{1\over \sqrt{2}}
\ve^{ijk}d^{\mf{abc}}Q_{k\mf c}
$$
\be
\label{e8com}
[Q^{i\mf a},\,Q_{j\mf b}]\,=
\,\delta^i_j (T_{\mf A})^{\mf a}{}_{\mf b} X_{\mf A}
\,-\,{i\over 2}\,\delta^{\mf a}_{\mf b} (\lambda_\Sigma)^i{}_j Y_\Sigma
\ee
From here the $E_8$ structural constants 
$[{\mb T}_{\mb A},{\mb T}_{\mb B}]={\mb C}^{\mb C}{}_{\mb{AB}}
{\mb T}_{\mb C}$ can be found.

The Killing tensor 
${\mb G}_{\mb{AB}}={\mb C}^{\mb C}{}_{\mb{AD}}{\mb C}^{\mb D}{}_{\mb{BC}}$
has the nonzero components:
\be
\label{e8kil}
\mb{G}_{\mf{AB}}\,=\,-\,30\,\delta_{\mf{AB}}\;,\qquad
\mb{G}_{\Sigma\Lambda}\,=\,-\,30\,\delta_{\Sigma\Lambda}\;,\qquad
\mb{G}_{i\mf a}{}^{j\mf b}\,=\,\mb{G}^{j\mf b}{}_{i\mf a}\,=
\,30\,\delta_i^j\delta_{\mf a}^{\mf b}
\ee

Representation 248 is real; so its element 
$A=A^{\mb A}{\mb T}_{\mb A}$ is antihermitean $A^+=-A$  if
$A_{i\mf a}=-(A^{i\mf a})^*$.

\end{document}